\documentclass[%
aip,
 jmp,%
 amsmath,amssymb,
preprint,%
]{revtex4-1}

\usepackage{graphicx}

\usepackage{dcolumn}%
\usepackage{bm}%

\def\ltsima{$\; \buildrel < \over \sim \;$}
\def\gtsima{$\; \buildrel > \over \sim \;$}
\def\simlt{\lower.5ex\hbox{\ltsima}}
\def\simgt{\lower.5ex\hbox{\gtsima}}

\usepackage{color}

\begin{document}
\preprint{AIP/123-QED}

\title{Hyperbolic method to explore multiplicity flow solutions in a four-sided lid-driven cavity}%
\author{ Hubert BATY}
\email{hubert.baty@unistra.fr}

\affiliation{Observatoire Astronomique de Strasbourg, Universit\'e de Strasbourg, CNRS, UMR 7550, 11 rue de l'Universit\'e, F-67000 Strasbourg, France}
\date{\today}%

\begin{abstract}
In this study, the hyperbolic method is adopted to explore the flow field states of incompressible flow in a four-sided lid-driven square cavity.
In particular, we focus on the flow bifurcation obtained at the critical Reynolds number $R_e \simeq 130$. In the hyperbolic method, the diffusive term is
transformed into an hyperbolic one by introducing a diffusion flux term, which is the solution of an additional equation.
A classical Riemann-like solver with a finite-volume discretization is thus employed for the full flux (splitted into advective and diffusive parts), in order to solve the steady-state incompressible Navier-Stokes equation.
The incompressibility of the flow is treated via the artificial pseudo-compressibility method.
It is shown that our numerical code is able to detect the bifurcation, by the analysis of the residual term relaxation during the pseudo-time iteration procedure.
Moreover, depending on the combination choice of slope limiters for the two spatial directions, our method is able to select the first or the second stable solution
among the double flow field state obtained when the Reynolds number is higher than the critical value that is estimated to be $129.4$ in our study.

\end{abstract}


\maketitle

\clearpage

\section{Introduction}

The two-dimensional flow driven by boundaries moving inside a cavity is considered to be a classical problem
of computational fluid mechanics. In this work, as in many studies, we consider a simple square cavity. This problem
display many interesting solutions, like vortex structures and multiple flow solutions [1-4].

The different studies have focused on cavity driven by single plate, two parallel plates, two non-facing plates, or four-sided plates.
In the latter configuration, the upper plate moves rightwards, the lower plate moves leftwards, the left plate moves downwards, and
the right plate moves upwards. The four plates are also considered to move at the same velocity in the present work. A remarkable result is the existence of
a bifurcation for a particular Reynolds number ($R_e$) value close to $R_{c} = 130$. Indeed, the steady-state flow changes from a single solution for
a Reynolds number value lower than $R_{c}$ to a double solution for a higher Reynolds number. The exact value obtained for $R_c$ in the different previous
numerical studies, varies between $129$ and $130.4$.

In the present study, we propose to revisit this problem by using the hyperbolic method. The hyperbolization idea is to transform the elliptic
diffusive term into an hyperbolic one by introducing the diffusive flux as an additional variable. This new variable is thus solution of
an additional hyperbolic equation. The advantages of this method are, (1) the discretization used for the non-dissipative (inviscid) part of the equations
can be directly applicable, (2) a speed-up factor of $O (1/h)$ is obtained ($h$ being the typical mesh spacing), (3) the diffusive terms are computed to the
same order of accuracy as the main solution. 
We also consider the two dimensional incompressible Navier-Stokes equation, with the incompressibility of the flow being ensured via the
artificial pseudo-compressibility technique. The hyperbolic technique has been successfully applied to diffusion, advection-diffusion, Navier-Stokes, and magnetohydrodynamic
equations [5-9].

The paper is organized as follows. Section II is devoted to the model and numerical method. In section III, the validation of the  scheme and tests are presented for the single-sided cavity case.
The results for the four-sided lid-driven cavity are presented in section IV, with a particular emphasis on the bifurcation of the steady-state flow solution.
Finally, conclusions are drawn in section V.

\section{Model and numerical method}
\label{setup}

\subsection{Navier-Stokes model}
The time dependent Navier-Stokes (NS) equation is generally written as,
\begin{equation}
    \frac{ \partial  \textbf{v}}   {\partial t } + (\textbf{v} \cdot  \nabla) \textbf{v} + \nabla P -  \nu  \nabla^2 \textbf{v}  = 0 ,
\end{equation}
for the velocity flow $\textbf{v}$, where $P$ is the kinematic thermal pressure (i.e. thermal pressure divided by the fluid density), and $\nu$ is the kinematic viscosity coefficient.
Thus, the steady-state (i.e. $ \frac{ \partial}{\partial t} \equiv 0 $, $t$ being the time) Navier-Stokes equation follows,
\begin{equation}
 (\textbf{v} \cdot  \nabla) \textbf{v} + \nabla P -  \nu  \nabla^2 \textbf{v}  = 0 .
\end{equation}
The incompressibility property must be also imposed via,
\begin{equation}
 \nabla  \cdot  \bf v = 0 .
\end{equation}
For an homogeneous viscosity, an equivalent conservative form of equations 2-3 is,
\begin{equation}
 \text {div} (\textbf{v}  \otimes \textbf{v}  + P  \bar{\bar I} ) -  \text {div}  ( \nu \bar{\bar g}) = 0,
\end{equation}
with $\bar{\bar g}  = ( \text {grad}  \textbf{v})$ being the velocity gradient tensor, and where
 $\bar{\bar I}$ is the identity rank-$2$ tensor.
The last equation makes appear two flux tensors, the advective flux modified by the pressure effect $\bar{\bar F}_a = (\textbf{v}  \otimes \textbf{v}  + P  \bar{\bar I} )$ which can be also
called the inviscid flux below, and the viscous stress tensor $\bar{\bar F}_v = - \nu \bar{\bar g}$. The full flux is thus $\bar{\bar F} = \bar{\bar F}_a + \bar{\bar F}_v$.

\subsection{Artificial compressibility method}
The previous incompressible Navier-Stokes equation can be solved by introducing a pseudo-time parameter $\tau$ in the following compressible equations,
\begin{equation}
     \frac{ \partial  P}   {\partial \tau }   +   \text {div} (c^2 \textbf{v}) = 0,
\end{equation}
and,
\begin{equation}
     \frac{ \partial  \textbf{v}}   {\partial \tau }   +     \text {div} (\textbf{v}  \otimes \textbf{v}  + P \bar{\bar I}   ) -  \text {div}  (\nu \bar{\bar g})  = 0,
\end{equation}
with $c^2$ the square value of $c$, an artificial sound speed. The value of $c$ is arbitrarily chosen in order ensure that the steady-state solution of equations 5-6  (i.e. $ \frac{ \partial}{\partial \tau} \equiv 0 $) is
equivalent to the originel incompressible Navier-Stokes equation. In practical, the value of $c$ is chosen to be one in this work.
This pseudo-incompressible form is said to be equivalent to the incompressible NS equations in the steady-state [10].

\subsection{The hyperbolic method}

It is possible to introduce another differential equation for the stress tensor as,
\begin{equation}
    T_r  \frac{ \partial \bar{\bar g} }   {\partial \tau }   -   \text {grad}  \textbf{v}  = - \bar{\bar g }  ,
    \end{equation}
where $T_r$ is the relaxation parameter defined by $T_r = L_r^2/\nu$, with $L_r$ a length-scale parameter. The value of $L_r$ value is determined
in order to achieve an optimal fast convergence [5-7].
In this way, we have $ \bar{\bar g } =  \text {grad}  \textbf{v} $ as expected in the steady-state.

In two dimensions, the system 5-7 can be expressed in the following compact form,
\begin{equation}
     \frac{ \partial  \textbf{U}}   {\partial \tau  } +    \frac{ \partial  \textbf{F}}   {\partial x }   +   \frac{ \partial  \textbf{G}}   {\partial y }   =    \textbf{S}  ,
    \end{equation}
with $ \textbf{S}$ being a source term containing the $\bar{\bar g }$ tensor, and where $\textbf{F}$ and $\textbf{G}$ represent $x$-directed and $y$-directed full flux respectively.

In order to fix the ideas, we can write a simpler one-dimensional 1D ($x$ dependent) version below, where the fluid velocity flow is given by $u(x)$. 
Then, the set of equations becomes,
\begin{equation}
       \frac{ \partial  P}   {\partial \tau }   +   \frac{ \partial  (c^2 u)}   { \partial x }   = 0,
    \end{equation}
\begin{equation}
       \frac{ \partial  u}   {\partial \tau }   +   \frac{ \partial  (u^2 + P - \nu g)}   { \partial x }   = 0,
    \end{equation}
\begin{equation}
       \frac{ \partial  g}   {\partial \tau }   -    \frac{ 1}   { T_r}    \frac{ \partial  u}   { \partial x }   = -  \frac{ g}   { T_r} .
    \end{equation}
In this one dimensional version, $\textbf{U}$ is a column vector containing three components, the pressure $P$, the velocity $u$, and $g$ the gradient of $u$. The corresponding flux $\textbf{F}$ has
three components that are, $c^2 u$, $u^2 + P - \nu g$, and $-u/T_r$. Finally, the source term $\textbf{S}$ has three components, that are zero for the two first ones and $- g/T_r$ for the third one.
It is important for the following numerical implementation to determine the eigenvalues of the inviscid flux Jacobian (i.e. ignoring the viscous terms), that are $ \lambda_ {\pm}= u \pm \sqrt {u^2 + c^2}$. 

\subsection{Discretization and implementation in one dimension}

We consider the 1D version below, as it is simpler to write. For a one-dimensional grid of $N$ nodes with a uniform spacing $h$, the solution can be evaluated at the nodes denoted by $x_j$, 
$j = 1,2,3,...,N$. Our finite-volume cell-centered discretization is thus,
\begin{equation}
     \frac{ \partial  \textbf U_j}   {\partial \tau  } =   - \frac{ 1}   {h} [ \textbf F_{ j+1/2}  - \textbf F_{ j-1/2}] + \frac{ 1} {h} \int_{I_j}  \textbf S \, \mathrm{d}x \ ,
         \end{equation}
where $\textbf U_j$ denotes the solution value at the node $j$, $\textbf F_{ j+1/2}$ is an interface flux to be defined (see below), and $I_j = [x_{ j-1/2}, x_{ j+1/2}]$ is a dual cell.

Steady state solutions of the previous system can be obtained by using a simply pseudo-time explicit iteration,
\begin{equation}
       \textbf U_j^{ n+1}   = \textbf U_j^{ n} -  \Delta \tau Res_j^n ,
         \end{equation}
where $n$ is the iteration counter, $\Delta \tau$ is the pseudo-time step, and $- Res_j^n$ is the residual right-hand side of equation (12). The latter residual term is
required to vanish (up to a pre-defined given accuracy) when a steady state solution is obtained.

The numerical fluxes can be estimated using the standard upwind formula,
\begin{equation}
      \textbf F_{ j+1/2} = \frac{ 1} {2} ( \textbf F_L +  \textbf F_R ) -  \frac{ 1} {2}  c_M ( \textbf U_R - \textbf U_L ),
         \end{equation}
where the subscripts L and R stand for the left- and right--hand sides of the cell interface situated at $x_{ j+1/2}$ respectively. The first term is computed from an
average value of the two fluxes $ \textbf F_L = \textbf F( \textbf  U_L)$ and $ \textbf F_R = \textbf F( \textbf  U_R)$. Instead of using a full flux directly, the 
full flux can be separated into two parts (inviscid and viscous parts), for which the above numerical formula can be used separately (see paper [9]).
The two corresponding flux Jacobians have the following eigenvalues $ \lambda_ {\pm}$ and $\pm \sqrt {\nu /T_r}$ for the inviscid and viscous terms respectively. As a consequence,
For the inviscid flux, $c_M$ is taken as the maximum absolute value of the two eigenvalues $ \lambda_ {\pm} = u \pm \sqrt {u^2 + c^2}$ and represents a maximum inviscid wave speed. For the diffusive flux, $c_M$ 
is simply $ \sqrt {\nu /T_r} = \nu/L_r$, and can be interpreted as an artificial viscous wave speed.
The source term in equation (12) is computed using a simple point integration approximation, that is shown to not degrade the accuracy of the method.

In the explicit pseudo-time iteration, the value of the increment in the pseudo-time variable $\tau$ is limited by a criterion,
\begin{equation}
\Delta \tau = CFL \times Min \left [ \frac{h}   {c_M + \nu/L_r}  \right ],
\end{equation}
where CFL is the Courant-Friedrichs-Lewy number less than or equal to one, and where we have simply added the maximum wave speed to the dissipative wave speed
in the denominator.

\subsection{Left/right interfaces interpolation}
The left and right values can be evaluated by a linear extrapolation from the cell centre, as for example for the $u$ solution of the 1D discretization,
\begin{equation}
      u_L = u_j + \frac{ 1} {2} h  \nabla u_j               ,      \ \  \  \  \            u_R = u_j -  \frac{ 1} {2} h  \nabla u_j   ,
  \end{equation}
where $\nabla u_j $ is the gradient of $u_j$. A simple and economic method is to use $g_j$ to evaluate the gradient term, as $g$ is the spatial derivative of $u$ in the steady state.
This is a second order scheme that has been previously shown to converge rapidly towards the steady state. Moreover, some classical limited reconstructions options can be easily
added to the scheme, that are the superbee, Koren, monotized-central, or minmod limiters.

\subsection{Discretization and implementation in two dimensions}

The previous one dimensional model can be also easily discretized in two dimensions. The resulting set of hyperbolic equations can be obtained
from the work done by Nishikawa (see [8]), and are not detailed in this paper. Our algorithm only differs by the splitting procedure of the full Flux (and the
corresponding Jacobian) into inviscid and viscous parts.

\begin{figure}[!ht]
\centering
 \includegraphics[scale=0.12]{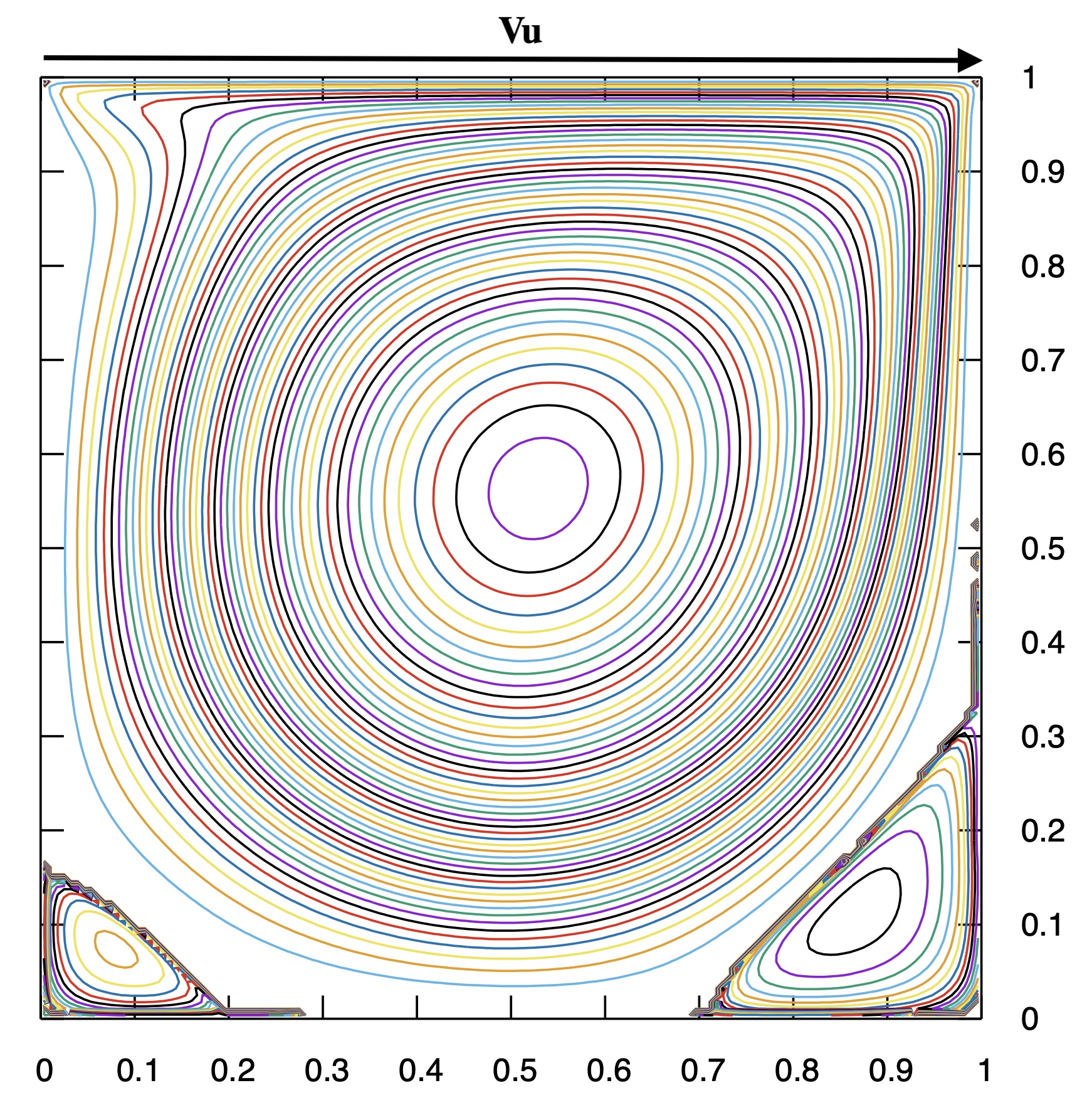}
  \caption{Colored iso-contours of a stream function (deduced from the velocity field) showing the flow stream lines for a one-sided cavity flow, obtained for a run using $R_e = 1000$.
  Negative values in the two lower corners indicate reversal of the vortices rotation compared to the rest of the domain.
    }
\label{fig1}
\end{figure}

\begin{figure}[!ht]
\centering
 \includegraphics[scale=0.12]{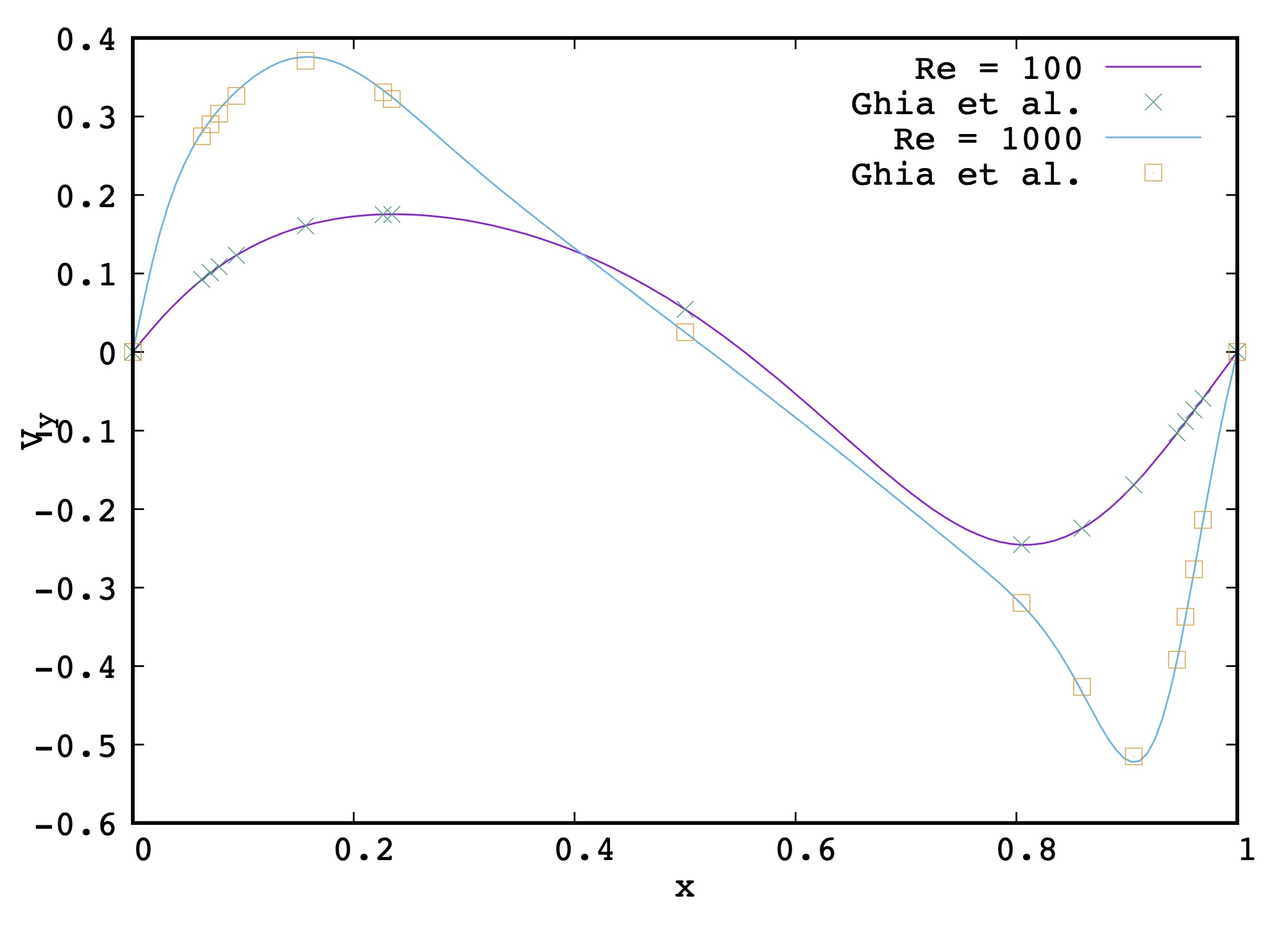}
  \caption{$V_y (x)$ profile obtained for $y = 1/2$, in two runs using $R_e = 100$ and $R_e = 1000$. A few selected values (see points)
  extracted from Ghia et al. (1982) are also shown for these two Reynolds number values.
   }
\label{fig2}
\end{figure}

\section{Application to the lid-driven cavity - one sided case}

As a test case, we first consider the one sided square cavity, where only the upper plate is moving with a velocity $V_x = V_u = 1$ (see figure 1).
Using the numerical scheme described above in two dimensions with a spatial domain $[0 :1]^2$, the resulting solution for the flow field lines is plotted
in figure 1 for a Reynolds number $R_e = 1/\nu = 1000$. Note that a minmod limiter is chosen for this run. A grid of $130 \times 130$ cells is also employed.
We have also found that taking $L_r = \frac{ 1} {20  \pi} $, allows an optimal and fast convergence of our results towards steady-state solutions.
The stream lines solution plotted in figure 1 is similar to the solution expected from previous studies [1-4]. Indeed, for this relatively high $R_e$ value, the two counter rotating
 vortices are barely visible at the two lower cavity corners.

In order to check the validity and convergence of our results, we have plotted in figure 2 the cross cut along an horizonthal line at $y = 1/2$ of
the $V_y$ velocity flow component as a function of $x$ for two Reynolds number values, $R_e = 1000$ and $R_e = 100$. Note that an optimal value of
$L_r = \frac{ 1} {20  \pi}$ is chosen in order to ensure a fast convergence of the results. Our results are compared with the results obtained by Ghia et al. (1982)
that is the usual numerical solution taken as the reference in the literature [11]. As one can see in figure 2, the agreement is very good.

\section{Application to the lid-driven cavity - four sided case}

We focus now on the four-sided square cavity case, where the velocity value imposed at each plate is taken to be one. In figure 3, one can see the solution obtained for $R_e = 100$, that is in agreement
with the expected solution shown in the literature. In order to examine the convergence of the solution towards the steady state, the residual
value taken on $V_x$ is plotted as a function of the iteration step in figure 4. Note that, even if a very good approximate solution for 
the true numerical steady state is already obtained after only $10 000$ iterations (the residual being $10^{-9}$), it can be even further ameliorated until $40 000$ iterations.

\begin{figure}[!ht]
\centering
 \includegraphics[scale=0.12]{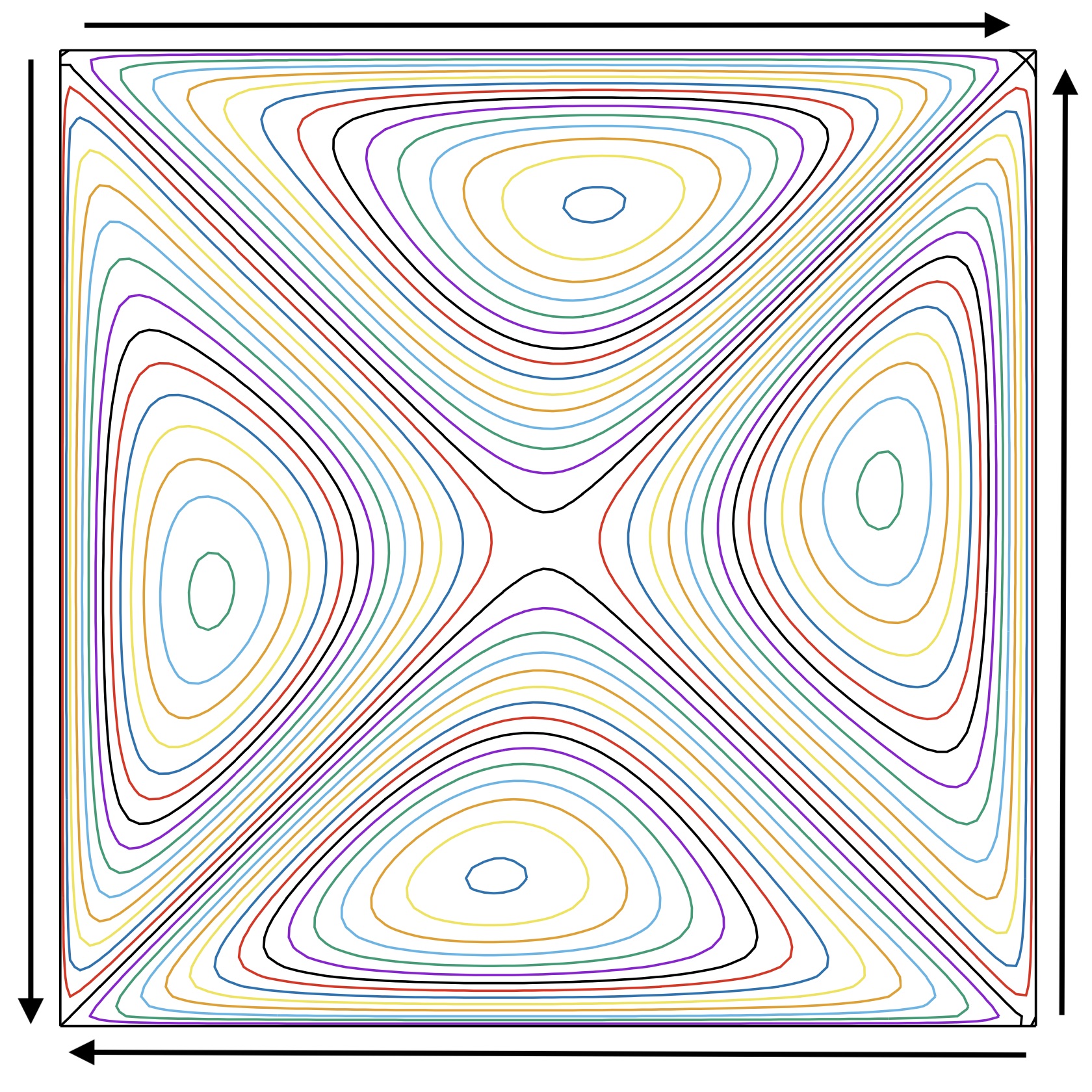}
  \caption{Colored isocontours of a stream function (deduced from the velocity field) showing the flow stream lines for a four-sided cavity case, obtained for a run using $R_e = 100$.
  The direction imposed on each flow plate is indicated.
    }
\label{fig3}
\end{figure}

\begin{figure}[!ht]
\centering
 \includegraphics[scale=0.12]{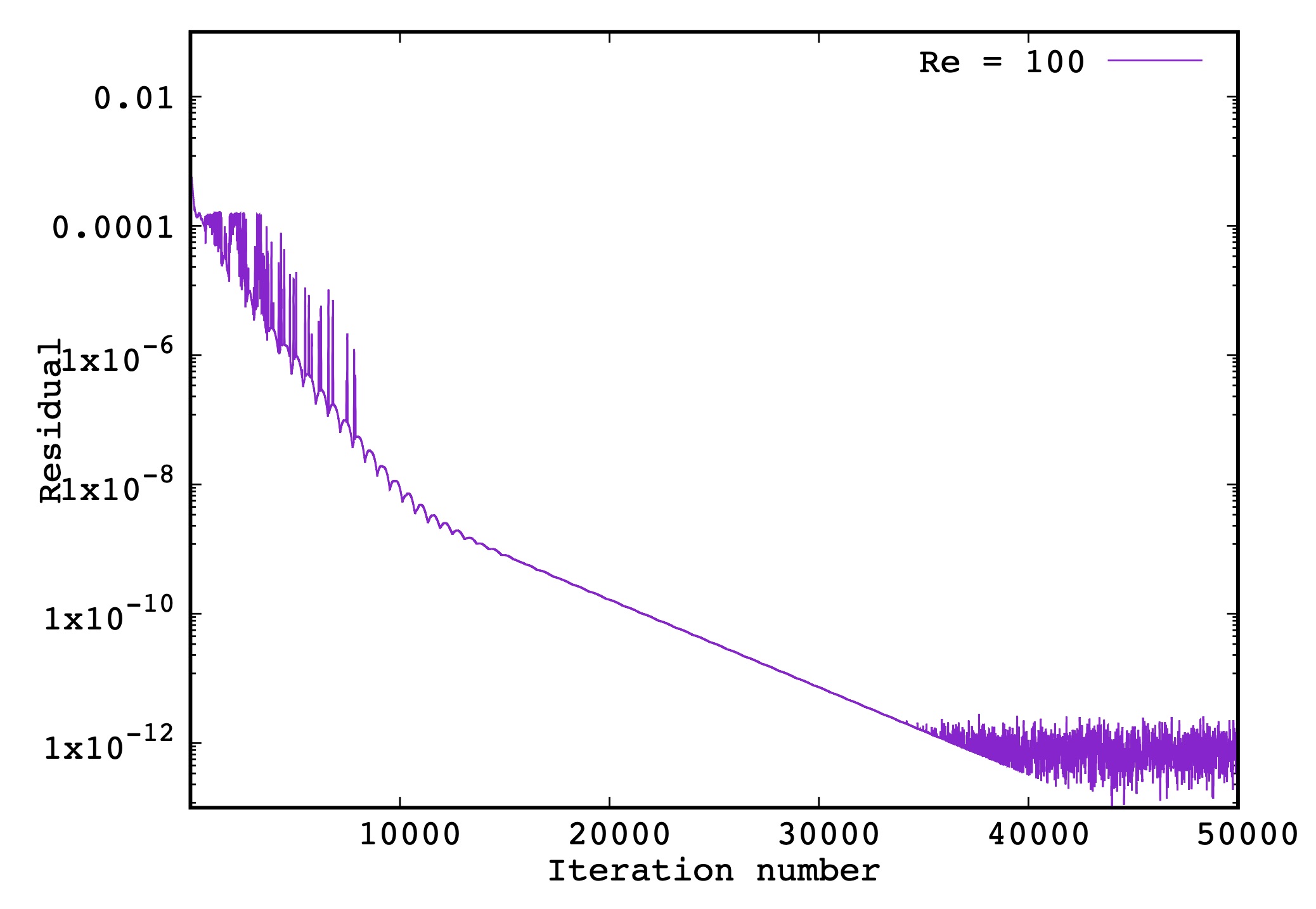}
  \caption{Residual term taken on one velocity flow component (that is $V_x$) as a function of the iteration counter, for a run of a four-sided cavity flow using $R_e = 100$ (see previous figure).
   }
\label{fig4}
\end{figure}

Now, we explore different Reynolds number values by increasing it over the previous value of $R_e =100$. The results of the residual term evolution as a function of the iteration counter are plotted
in figure 5 for $4$ values of the Reynolds number ranging between $100$ and $160$. A change in the convergence behavior is clearly visible between $R_e = 120$ and $R_e = 140$. Indeed, contrary
to the monotonic decrease of the residual obtained for the two lowest $R_e$ values, a clearly non monotonic variation is seen to precede the final convergence phase for the two highest Reynolds numbers.
We have checked that it is in correspondance to the existence of two possible stable steady-state flow solutions, as shown in figure 6 for $R_e = 140$. The two previous solutions only differ by the use of different
combination of slope limiters in $x$ and $y$ directions.
Consequently, we take this change of convergence behavior as a good diagnostic in order to determine the critical value of the Reynolds number $R_c$ for the bifurcation. We
precisely get $R_c = 129.4$ with our procedure.

\begin{figure}[!ht]
\centering
 \includegraphics[scale=0.3]{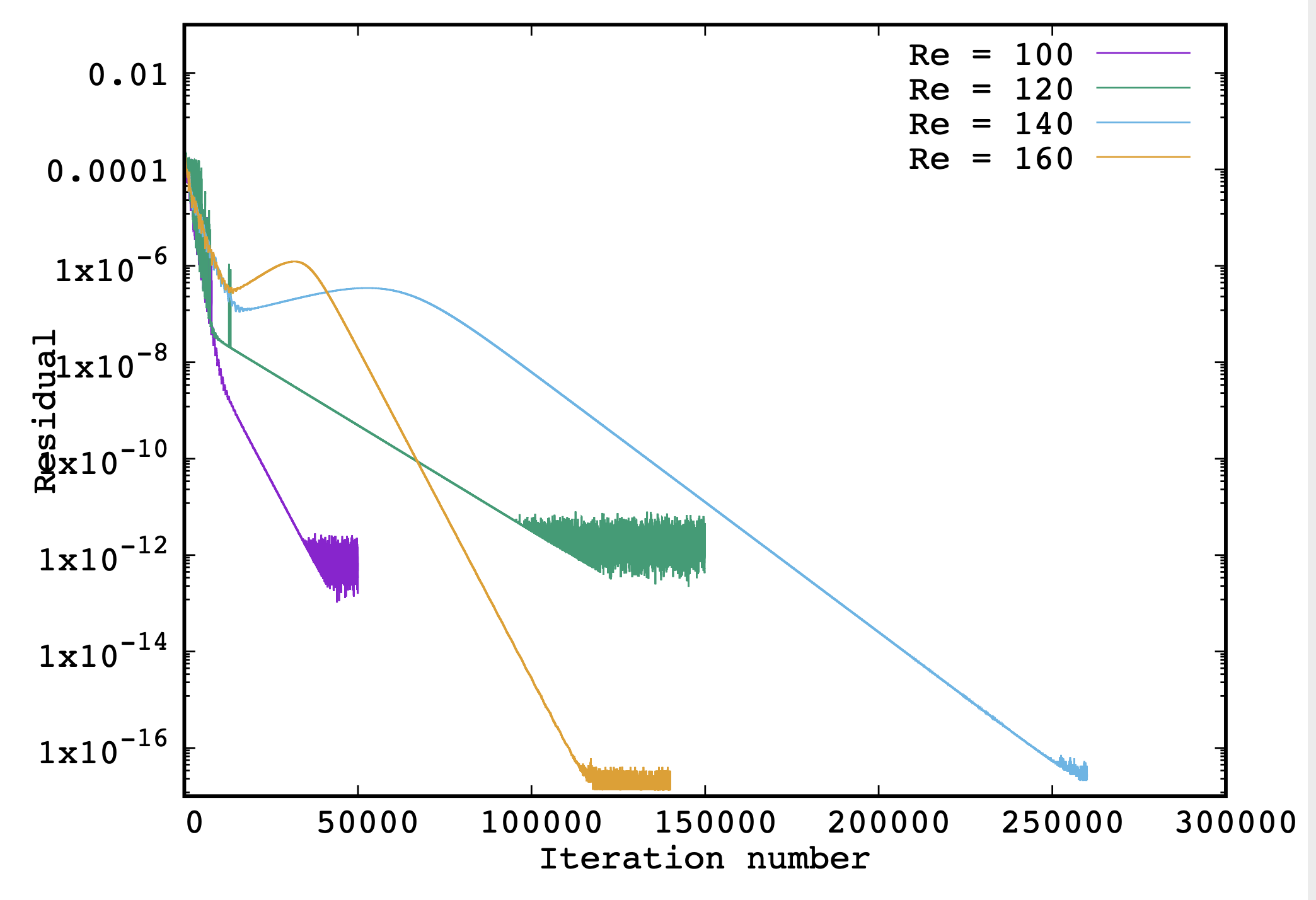}
  \caption{Residual term taken on one velocity flow component (that is $V_x$) as a function of the iteration counter, for runs of a four-sided cavity flow using $R_e = 100, 120, 140, 160$.
   }
\label{fig5}
\end{figure}

\begin{figure}[!ht]
\centering
 \includegraphics[scale=0.1]{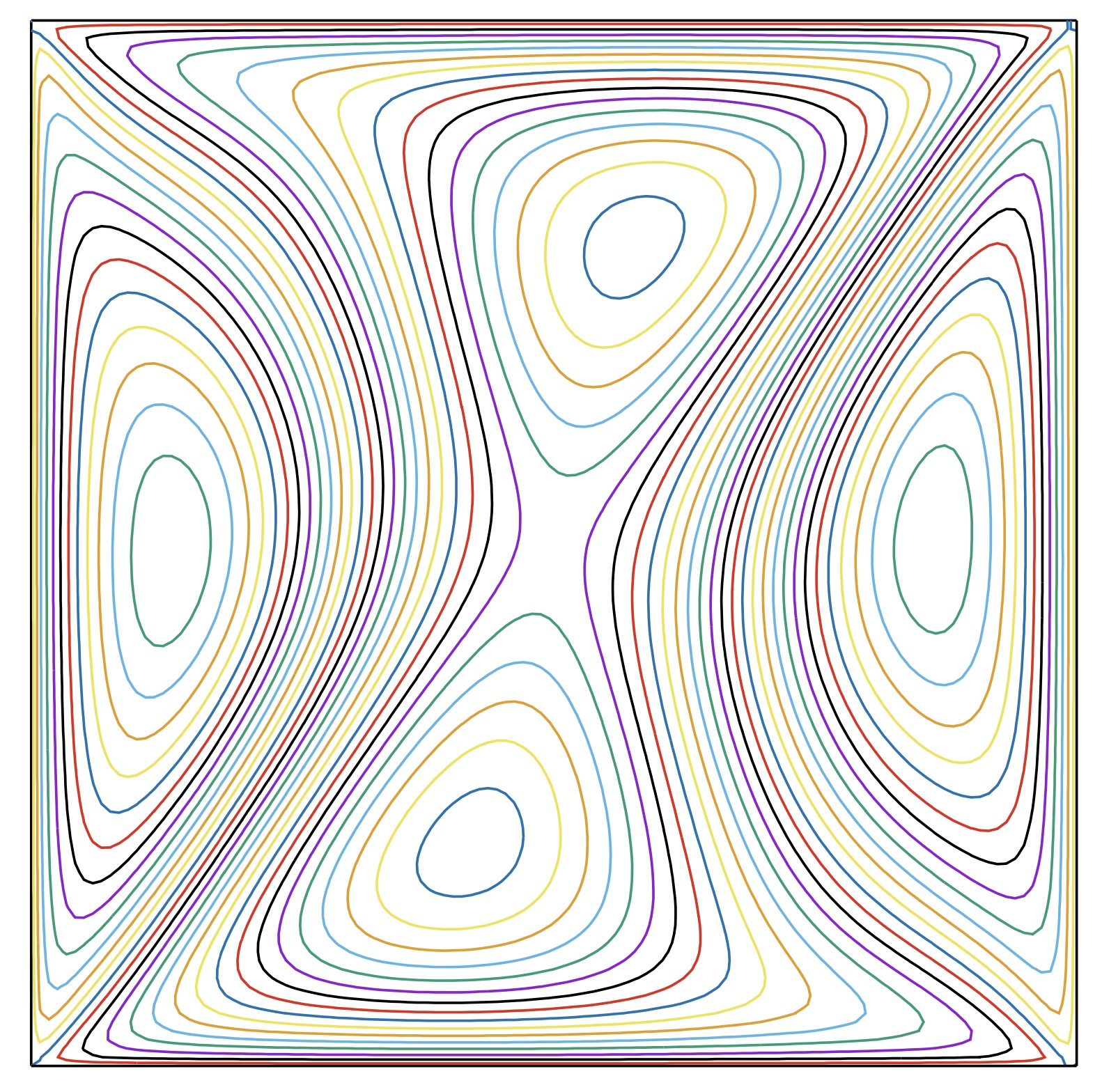}
 \includegraphics[scale=0.1]{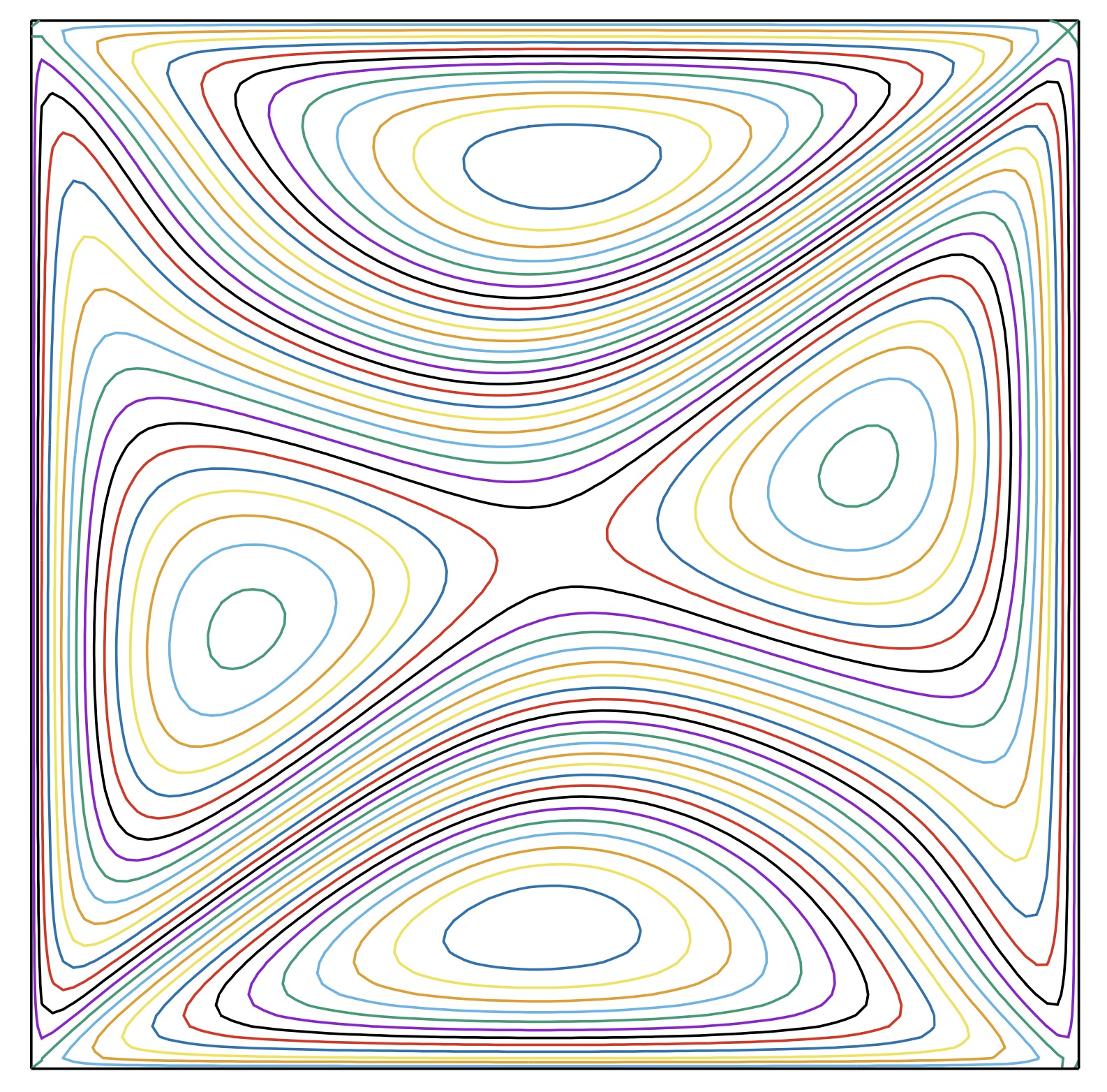}
  \caption{The two stable steady state flow solutions (stream lines) obtained in two runs of a four-sided cavity flow using $R_e = 140$. The two simulations only differ by the use of different slope limiters combination in $x$ and $y$
  directions.
   }
\label{fig6}
\end{figure}

\section{Conclusion}
In this study, we use the hyperbolic method in order to study steady state flow solutions of the four-sided lid-driven square cavity in two dimensions. A classical Riemann solver is used
to solve the incompressible Navier-Stokes equation with the artificial compressibility method. A simple second-order finite-volume discretization on rectangular grid using upwind
advective and dissipative fluxes is employed. Our results show a remarkable efficiency to determine the critical Reynolds number $R_c$ corresponding to a the bifurcation between
a single stable solution at low Reynolds number to a doubly stable one when the Reynolds number is increased. This is precisely obtained by the examination of the convergence
behavior of the residual terms during the explicit pseudo-time iteration. We get the critical value $R_c = 129.4$, which is in the range of values $[129 : 130.4]$ previously deduced in
the literature.

\end{document}